\documentclass[prl,aps,twocolumn,showpacs,preprintnumbers,amsmath,amssymb]{revtex4}
\usepackage{dcolumn}
\usepackage{bm}
\usepackage{amsmath,amssymb,amsfonts,latexsym,fancyhdr,graphicx}

\newcommand{\ket}[1]{\displaystyle{|#1\rangle}}
\newcommand{\bra}[1]{\displaystyle{\langle#1|}}
\newcommand{\al}{\alpha}
\newcommand{\Om}{\Omega}
\newcommand{\om}{\omega}

\newcommand{\g}{\gamma}
\newcommand{\si}{\sigma}
\newcommand{\la}{\lambda}

\begin{document}
\title{Sudden death and sudden birth of entanglement in common structured reservoirs}
\author{L. Mazzola,$^1$ S. Maniscalco,$^1$ J. Piilo,$^1$ K.\,-\,A. Suominen,$^1$ and B. M. Garraway$^2$}
\affiliation{$^1$Department of Physics and Astronomy, University of
Turku, FI-20014 Turun yliopisto, Finland\\$^2$Department of Physics
and Astronomy, University of Sussex, Falmer, Brighton, BN1 9QH,
United Kingdom}
\email{sabrina.maniscalco@utu.fi}

\begin{abstract}
We study the exact entanglement dynamics of two qubits in a common
structured reservoir. We demonstrate that, for certain classes of
entangled states, entanglement sudden death occurs, while for
certain initially factorized states, entanglement sudden birth takes
place. The backaction of the non-Markovian reservoir is responsible
for revivals of entanglement after sudden death has occurred, and
also for periods of disentanglement following entanglement sudden
birth.
\end{abstract}

\maketitle

Entanglement is one of the most intriguing features of quantum
mechanics~\cite{Harbook}: hence understanding its properties and
dynamics is of paramount relevance for a number of applications in
modern physics. Furthermore, quantum cryptography, quantum information processing
and quantum measurement are examples of branches of physics in
which entanglement plays an essential role~\cite{Nielsen}. Quantum
properties, however, are very fragile: realistic quantum systems are
not closed, and due to the interaction with the environment, their
entanglement and coherence can be irretrievably lost. On the other
hand, recent works~\cite{Bellomo,Zeno,Etrapping} have shown that
entanglement can actually revive, or be preserved, using the quantum
Zeno effect, or it can even be trapped. Such effects arise when the
system of interest interacts with non-Markovian reservoirs. These
reservoirs are characterized by structured spectral
distributions~\cite{Breuer}.

In this Letter we propose an exactly solvable model for the time
evolution of two qubits interacting with a common structured
reservoir. We investigate the entanglement dynamics for initially
entangled states of the qubits, and also for initially factorized
states. We identify the effects of the non-Markovian backaction of
the reservoir during the disentangling process and in the
entanglement birth phenomenon. In particular we prove that, due to
the reservoir memory effect, entanglement sudden death (ESD) is
enhanced as compared to the Markovian case. Contrary to a previous
study on ESD in common environments~\cite{Leon}, our approach is
exact and it does not rely on either the Born or the Markov
approximation.

The entanglement dynamics of two qubits interacting with two
\emph{independent} Lorentzian reservoirs  are known~\cite{Bellomo}. In this
Letter we focus on the \emph{common} reservoir scenario and compare it with
the independent reservoirs case. We note that, already in the
Markovian case, the entanglement dynamics in common and independent
reservoirs present striking differences. While two qubits
interacting with two independent reservoirs can disentangle
completely and permanently in a finite time~\cite{Yu,Almeida}, in a
common reservoir entanglement can disappear for a finite time and
then reappear again~\cite{Ficek}. This is due to the fact that
common reservoirs tend to create entanglement rather than destroy it
completely, since they indirectly couple the two qubits
\cite{Braun}. As we show in the following, even when the qubits are
initially prepared in a factorized state, the correlation created by
the environment can lead to a phenomenon which is known as the
entanglement sudden birth (ESB)~\cite{FicekESB}.

The analytic solution presented in this Letter is valid for a
general initial state of the two-qubit system, and includes states
with two excitations. The exact solution for the single excitation
case has been discussed in Ref.~\cite{Zeno}, where it was shown that
the initial entanglement never disappears for finite periods of
time, i.e., ESD never occurs. When two excitations are present, as
for the model discussed here, the derivation of the exact analytical
solution is much more complicated. We have proved, however, that by
using the pseudomode approach~\cite{Barry97} and establishing a
connection with a three-level ladder system~\cite{Barry3}, it is
still possible to solve the dynamics without performing any
approximations.

We consider a two-qubit system interacting with a common
zero-temperature bosonic reservoir. Our chosen specific system
consists of two two-level atoms interacting with the electromagnetic
field. The initial state of the multimode field is the vacuum state.
The Hamiltonian of such a system in the rotating wave approximation
is given by $H=H_{0}+H_{int}$, which, in the basis
$\{\ket{00},\ket{10},\ket{01},\ket{11}\}$, reads
\begin{equation}
   H_{0}=\om_{0}(\si_{+}^{A}\si_{-}^{A}+\si_{+}^{B}\si_{-}^{B})
   +\sum_{k}\om_{k}a_{k}^{\dag}a_{k},\label{H0bare}
\end{equation}
\begin{equation}
   H_{int}=(\si_{+}^{A}+\si_{+}^{B})\sum_{k}g_{k}a_{k}+\mathrm{h.c}.\label{Hintbare}
\end{equation}
Here, $\si_{\pm}^{A}$ and $\si_{\pm}^{B}$ are, respectively, the
Pauli raising and lowering operators for the atoms A and B,
$\om_{0}$ is the Bohr frequency of the two atoms, $a_{k}$ and
$a_{k}^{\dagger}$ are the annihilation and creation operators for
the field mode $k$, and mode $k$ is characterized by the frequency
$\om_{k}$ and the coupling constant $g_{k}$.
For the sake of simplicity, in the following we assume that the two
atoms interact resonantly with a Lorentzian structured reservoir,
such as, e.g., the electromagnetic field inside a lossy resonator.

Since the atoms are identical and equally coupled to the reservoir,
the dynamics of the two qubits can be effectively described by a
four state system in which three states are coupled to the vacuum in
a ladder configuration, and one state is completely decoupled from
the other states and from the field.
This can be shown by writing the Hamiltonian in the basis
$\{\ket{0}=\ket{00},\ \ket{+}=(\ket{10}+\ket{01})/\sqrt{2},
\ket{-}=(\ket{10}-\ket{01})/\sqrt{2},\ \ket{2}=\ket{11}\}$ so that
\begin{equation}
   \label{H0dressed}
   H_{0}=2\om_{0}\ket{2}\bra{2}+\om_{0}(\ket{+}\bra{+}+\ket{-}\bra{-})
   +\sum_{k}\om_{k}a_{k}^{\dag}a_{k},
\end{equation}
\begin{equation}
   \label{Hintdressed}
   H_{int}=\sum_{k}\sqrt{2}g_{k}a_{k}(\ket{+}\bra{0}+\ket{2}\bra{+})+\mathrm{h.c.}
\end{equation}
The $\ket{+}$ and $\ket{-}$ states are, respectively, the
superradiant and subradiant states. From Eqs.~\eqref{H0dressed}
and~\eqref{Hintdressed} it is apparent that the subradiant state
does not decay. On the contrary, the superradiant state is coupled
to states $\ket{0}$ and $\ket{2}$ via the electromagnetic field.

The total Hamiltonian, given by Eqs.~\eqref{H0dressed}
and~\eqref{Hintdressed}, actually consists of two parts, a part
describing the free dynamics of the $\ket{-}$ state, and the
remaining terms describing a three-state ladder system
$\{\ket{0},\ket{+},\ket{2}\}$  with the transitions
$\ket{0}\leftrightarrow\ket{+}$ and $\ket{+}\leftrightarrow\ket{2}$
having the same frequencies and identically coupled with the common
bosonic reservoir.

It is a mathematically and computationally demanding task to solve
numerically the infinite set of differential equations for the
complex amplitudes appearing in the state vector of the total
system. However, having in mind Eqs.~\eqref{H0dressed}
and~\eqref{Hintdressed}, we can greatly simplify the analytical
treatment by noting that the dynamics of the three-level ladder
system, characterized by transitions with equal frequencies and
identically coupled to the same Lorentzian structured reservoir, can
be exactly solved using the pseudomode approach~\cite{Barry97}. In
Ref.~\cite{quasimodes} this approach is extended to multiple
excitations.

The pseudomodes are auxiliary variables
defined from the properties of the spectral distribution, and they allow us
to derive a \emph{Markovian} master equation for the extended system
comprised of the system of interest and the pseudomodes. Such an exact
master equation describes the coherent interaction between the
system and the pseudomodes, and the latter, in turn, leak into
independent Markovian reservoirs.

The exact dynamics of the three-state system interacting with a
Lorentzian structured reservoir is contained in the following
pseudomode master equation
\begin{equation}\label{PseuME}
   \frac{\partial\rho}{\partial
   t}=-i[V,\rho]-\frac{\Gamma}{2}(a^{\dag}a\rho+\rho a^{\dag}a-2 a\rho a^{\dag}),
\end{equation}
where $\rho$ is the density matrix of the three\,-\,level system and
the pseudomode, and
\begin{equation}\label{V}
   V=\sqrt{2}\Om(a\ket{+}\bra{0}+a^{\dag}\ket{0}\bra{+}+a\ket{2}\bra{+}
   +a^{\dag}\ket{+}\bra{2}).
\end{equation}

Here, $a$ and $a^{\dag}$, $\Gamma$ and $\Om$ are, respectively, the
annihilation and creation operators, the pseudomode decay rate into
its Markovian reservoir, and the coupling constant of the pseudomode
to the ladder system. The pseudomode is associated with the
Lorentzian spectral distribution
\begin{equation}
   J(\omega)=\frac{\Omega^2}{\pi}\frac{\Gamma}{(\om-\om_{0})^2+(\Gamma/2)^2},
\end{equation}
where $\Gamma/2$ describes the frequency width of the spectrum and
is related to the reservoir correlation time. For small values of
$\Gamma$ the pseudomode can be associated with the real cavity mode
of frequency $\om_{0}$.

In order to find the dynamics of the three states we solve the
master equation in Eqs.~\eqref{PseuME} and~\eqref{V} and then trace
out the degrees of freedom of the pseudomode. Then we take into
account the $\ket{-}$ state and consider the density matrix for the
two atoms in the initial basis
$\{\ket{00},\ket{10},\ket{01},\ket{11}\}$.

Our aim is to investigate the effects of the non-Markovianity of the
reservoir on atomic entanglement dynamics. To quantify the entanglement
we use the Wootters concurrence~\cite{Wootters}, defined as
$C(t)=\textrm{max}\{0,\sqrt{\la_{1}}-\sqrt{\la_{2}}-\sqrt{\la_{3}}-\sqrt{\la_{4}}\}$,
where $\{\la_{i}\}$ are the eigenvalues of the matrix
$R=\rho(\si_{y}^{A}
\otimes\si_{y}^{B})\rho^{\ast}(\si_{y}^{A}\otimes\si_{y}^{B})$, with
$\rho^{\ast}$ denoting the complex conjugate of $\rho$ and
$\si_{y}^{A/B}$ are the Pauli matrices for atoms $A$ and $B$. This
quantity attains its maximum value 1 for maximally entangled states
and vanishes for separable states.

For initial states of the form
\begin{equation}\label{ESDinistate}
   \ket{\Psi(0)}=\al\ket{00}+e^{i\theta}(1-\al^2)^{1/2}\ket{11}
\end{equation}
the density matrix of the atomic system has an \lq\lq X\rq\rq\ form:
\begin{equation}
   \rho=\left(
      \begin{array}{cccc}
        a(t) & 0 & 0 & w(t) \\
        0 & b(t) & z(t) & 0 \\
        0 & z^{*}(t) & c(t) & 0 \\
        w^{*}(t) & 0 & 0 & d(t) \\
      \end{array}
   \right),
\end{equation}
with non-zero elements only along the main diagonal and
antidiagonal. Due to the structure of the differential equations for
the density matrix elements [see Eq.~\eqref{PseuME}], the \lq\lq
X\rq\rq\ form is preserved during the evolution. Then the concurrence
has a simple analytic expression
\begin{equation}\label{conc}
   C(t)=\mathrm{max}\{0,C_{1}(t),C_{2}(t)\},
\end{equation}
where
\begin{equation}\begin{split}\label{C1C2}
   C_{1}(t)&=2|w(t)|-2\sqrt{b(t) c(t)},\\
   C_{2}(t)&=2|z(t)|-2\sqrt{a(t) d(t)}.
\end{split}\end{equation}

\begin{figure}[!]
\begin{center}
\includegraphics[width=8.6cm]{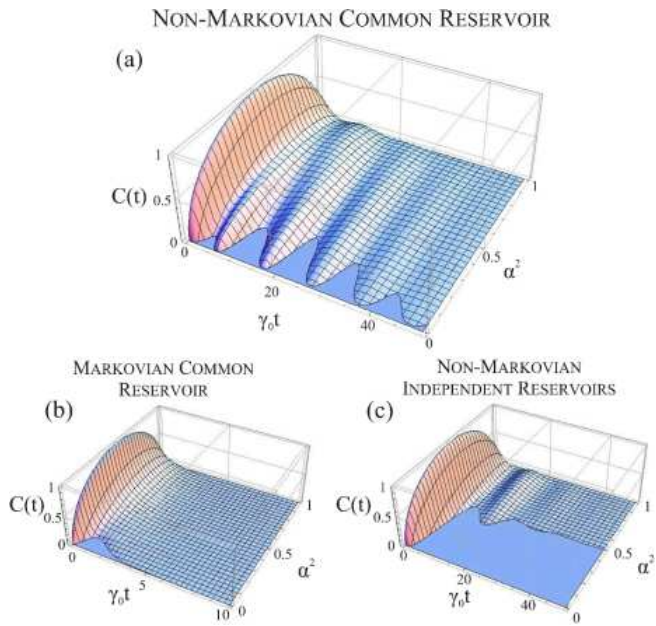}
\end{center}
\caption{(Color online) (a) Concurrence as a function of scaled time
and $\al^{2}$ for two atoms prepared in the
state~\eqref{ESDinistate} and interacting with a common Lorentzian
structured reservoir. For comparison we show also the equivalent
Markovian case (b) and the non-Markovian independent reservoirs
situation (c).}\label{fig:compESD}
\end{figure}

We plot the time evolution of concurrence as a function of both the
parameter $\al^{2}$ and the dimensionless quantity $\g_{0}t$,
where the parameter $\g_{0}=4\Om^{2}/\Gamma$ is the Markovian decay
rate of the atoms, i.e., the inverse of the atomic relaxation time
in the Markovian limit~\cite{Breuer}. The parameters $\Gamma$ and
$\Om$ are also expressed in units of $\g_{0}$. We look at the
dynamics in the strong coupling regime, which is obtained when
$\Gamma/\Om<4$; in particular we choose $\Gamma=0.2\ \g_{0}$ and
$\Om=\sqrt{0.05}\ \g_{0}$, corresponding to $\Gamma/\Om=\sqrt{0.8}$.
These values are achievable experimentally, e.g.,
with circuit QED setups \cite{Sillanpaa}.

Figure~\ref{fig:compESD}(a) shows the behavior of concurrence for
two qubits prepared in the state~\eqref{ESDinistate}. Depending
on $\al^{2}$ different dynamical behavior is clearly
visible. For $\al^{2}\gtrsim 1/4$ the entanglement dynamics presents
damped oscillations. For $\al^{2}\lesssim 1/4$ finite periods of
complete disentanglement are followed by entanglement revivals.
Entanglement revivals are amplified for stronger non-Markovian
conditions, i.e., for smaller values of the ratio $\Gamma/\Omega$.
We have verified that when the ratio $\Gamma/\Omega$ is, e.g., ten
times smaller than the one used in Fig.~\ref{fig:compESD}, the
oscillations become much stronger in amplitude, the number of ESD
periods and revivals increases, and they last for a longer time. On
the other hand, when we increase $\Gamma/\Omega$, the ESD regions
shrink and the period of the entanglement oscillations becomes
smaller. We note in passing that entanglement oscillations also
occur when the qubits share just one excitation (as shown in
Ref.~\cite{Zeno}). In that case the memory depth of the
reservoir leads to entanglement oscillations, but ESD does not occur.

Entanglement sudden death and entanglement revivals, in the common
structured reservoir, are basically due to two combined and
intertwined effects: the backaction of the structured reservoir and
the reservoir-mediated interaction between the qubits. In order to
understand the role played by each of these effects we compare our
results with the cases of the common Markovian reservoir, and two
independent non-Markovian reservoirs.

In Fig.~\ref{fig:compESD}(b) we show the entanglement dynamics for
two qubits in a common Markovian reservoir. As explained in
Ref.~\cite{Ficek}, a period of complete disentanglement is followed
by a revival of entanglement, but no oscillations are present. Such
a revival is due to the action of the common reservoir which tends
to create quantum correlations between the qubits, providing an
effective coupling between them. A comparison between
Figs.~\ref{fig:compESD}(a) and~\ref{fig:compESD}(b) shows that the
feedback of information from the reservoir into the system,
characterizing non-Markovian dynamics, enhances the appearance of
ESD regions, since it tends to recreate the conditions that led to
the first ESD period.

Figure~\ref{fig:compESD}(c) shows the entanglement dynamics when the
qubits interact with two independent non-Markovian reservoirs. Three
different dynamical regions are clearly identified. Depending on the
initial state of the system, entanglement can die after a finite
time, or oscillate while going asymptotically to zero, or reappear
after an ESD period. The revival phenomena in this case stem from
the non-Markovian behavior of each single qubit interacting with its
own reservoir~\cite{Bellomo}. No revivals of entanglement are
present in the Markovian case~\cite{Yu}.

A comparison between the non-Markovian cases of
Fig.~\ref{fig:compESD}(a) and~\ref{fig:compESD}(c) reveals that, for
the same type of reservoir spectrum, ESD regions are much wider in
the independent reservoirs case than in the common reservoir case.
Since both cases take into account memory effects, this suggests
that the reservoir-mediated interaction between the qubits, in the
common reservoir scenario, effectively counters the fast
disappearance of entanglement.

Non-Markovian effects also influence strongly the dynamics for
initially factorized states, when the reservoir-mediated interaction
between the qubits leads to entanglement generation. As an example
we take the initial state
\begin{equation}\begin{split}\label{ESBinistate}
   \rho(0)&=(\al^{2}\ket{0_{A}}\bra{0_{A}}+(1-\al^{2})\ket{1_{A}}\bra{1_{A}})\\
   &\quad\otimes (\al^{2}\ket{0_{B}}\bra{0_{B}}+(1-\al^{2})\ket{1_{B}}\bra{1_{B}}).
   \end{split}
\end{equation}
In the basis $\{\ket{0},\ket{+},\ \ket{-},\ \ket{2}\}$ this state
takes the form
\begin{equation}\begin{split}\label{ESBinistate2}
   \rho(0)&=\al^{4}\ket{0}\bra{0}+(1-\al^{2})^2\ket{2}\bra{2})\\
   &\quad +\al^{2}(1-\al^{2})(\ket{+}\bra{+}+\ket{-}\bra{-}).
\end{split}\end{equation}
Again the density matrix has the \lq\lq X\rq\rq\ form so the
concurrence can be calculated from Eqs.~\eqref{conc}
and~\eqref{C1C2}. We evaluate the concurrence for the same
parameters of the spectral distribution used in
Fig.~\ref{fig:compESD}(a).

\begin{figure}[!]
\begin{center}
\includegraphics[width=8.6cm]{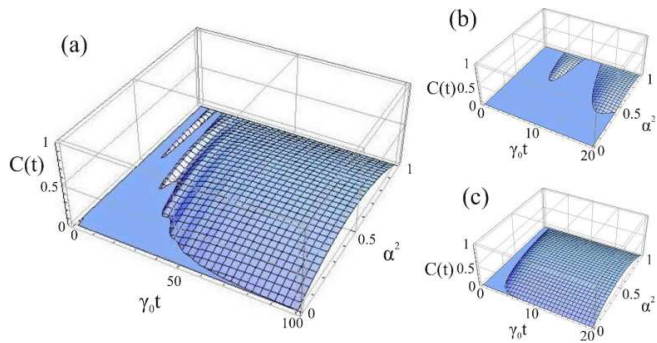}
\end{center}
\caption{(Color online) (a) Concurrence as a function of time and
$\al^{2}$ for two atoms prepared in the factorized
state~\eqref{ESBinistate} and interacting with a common Lorentzian
structured reservoir. For comparison we show the short-time region
of (a) in (b), and the Markovian common reservoir result in
(c).}\label{fig:ESB}
\end{figure}

Figure~\ref{fig:ESB} shows the evolution of concurrence when the
atoms are prepared in the factorized state~\eqref{ESBinistate}. We
note first that, for independent reservoirs, sudden birth of
entanglement does not appear, while in our common reservoir case,
ESB does occur, as clearly demonstrated in Fig.~\ref{fig:ESB}(a).
Moreover, in a structured reservoir such a phenomenon presents new
interesting features, compared to the Markovian case, as shown in
Figs.~\ref{fig:ESB}(b)-(c). To elaborate: in a Markovian reservoir,
entanglement sudden birth takes place at different times depending
on the parameter $\al^{2}$. The smaller the value of $\al^{2}$, the
longer the time taken for entanglement generation. In other words,
the ESB time monotonically decreases with $\al^{2}$. However, in the
non-Markovian case, ESB \emph{revivals} occur and complicate the
simple Markovian picture. In general, the reservoir memory prolongs
the initial disentanglement. In this case, therefore, the reservoir
backaction dominates over the tendency of the common reservoir to
create entanglement between the qubits. ESB periods and
disentanglement revivals become more frequent and numerous for
stronger non-Markovian conditions.

It is worth noting that a necessary condition for ESB is that
$\langle - \vert \rho \vert - \rangle \neq 0$. Indeed, the long-time
asymptotic value of concurrence is directly related to the
subradiant state component of the initial state~\cite{FicekESB}.

Whatever the value of $\al$ is, for $t \rightarrow \infty$ the
initial population of the three state ladder system will decay to
the $\ket{0}$ state, while the population of the subradiant
$\ket{-}$ state will be trapped. The asymptotic stationary state of
the system has the form
\begin{equation}
   \rho(t\rightarrow\infty)=(1-k)\ket{0}\bra{0}+k\ket{-}\bra{-},
\end{equation}
with $k=\al^{2}(1-\al^{2})$. For $\al^2\neq 0$, or $1$, this
state is not a factorized state. The stationary value of the
concurrence, calculated from the analytic solution, is $C(t
\rightarrow \infty)=k$.

In conclusion, we have presented a non-Markovian model describing
the exact entanglement dynamics of two qubits interacting with a
common structured reservoir. We have brought to light new
entanglement features for qubits prepared in both entangled and
factorized states. The non-Markovian nature of the reservoir
protracts the disentanglement process while enriching the revivals,
and at the same time it enhances the regions of ESD. The backaction
of the reservoir slows down the generation of entanglement and
further manifests itself in the appearance of periods of death and
resurrection. The reservoir-mediated interaction between the qubits
strikingly distinguishes the dynamics in a common reservoir from the
independent reservoirs case. Our predictions apply to cavity QED
experiments with trapped ions, and to circuit QED experiments. In
the first context, entanglement between two remotely located trapped
atomic ions has been recently demonstrated~\cite{Monroe} and
multiparticle-entangled states can be generated and fully
characterized via state tomography~\cite{Blatt}. In the second
context, field coupling and coherent quantum state storage between
two Josephson phase qubits has been achieved through a microwave
cavity on chip~\cite{Majer,Sillanpaa}. Due to the possibilities for
realizing strong coupling conditions between atoms and a high
finesse cavity~\cite{Wolfy,Wallraff} a deep understanding of the
non-Markovian dynamics is now indispensable.

The authors thank CIMO, the Academy of Finland (projects 108699,
115682 and 115982) and the Turku University Foundation for support.

\end{document}